\documentclass{pasj00}

\begin{document}
\SetRunningHead{S. Kameno et al.}{Plasma Torus around NGC~1052's Nucleus}
\Received{2000/12/27}
\Accepted{2001/02/09}
\SetVolumeData{2001}{53}{2}

\title{The Dense Plasma Torus Around the Nucleus of an Active Galaxy NGC~1052}

\author{%
    Seiji     \textsc{Kameno}\altaffilmark{1}
    Satoko    \textsc{Sawada-Satoh}\altaffilmark{2}
    Makoto    \textsc{Inoue}\altaffilmark{1}
    Zhi-Qiang \textsc{Shen}\altaffilmark{2}
    and
    Kiyoaki   \textsc{Wajima}\altaffilmark{2}}
  \altaffiltext{1}{National Astronomical Observatory, 2-21-1 Osawa, Mitaka, Tokyo 181-8588}
  \email{kameno@hotaka.mtk.nao.ac.jp}
  \altaffiltext{2}{The Institute of Space and Astronautical Science, 3-1-1 Yoshinodai, Sagamihara, Kanagawa 229-8510}
%


\maketitle

\begin{abstract}
A subparsec-scale dense plasma torus around an active galactic nucleus (AGN) is unveiled.
We report on very-long-baseline interferometry (VLBI) observations at 2.3, 8.4, and 15.4 GHz towards the active galaxy NGC 1052.
The convex spectra of the double-sided jets and the nucleus imply that synchrotron emission is obscured through free--free absorption (FFA) by the foreground cold dense plasma.
A trichromatic image was produced to illustrate the distribution of the FFA opacity.
We found a central condensation of the plasma which covers about 0.1 pc and 0.7 pc of the approaching and receding jets, respectively.
A simple explanation for the asymmetric distribution is the existence of a thick plasma torus perpendicular to the jets.
We also found an ambient FFA absorber, whose density profile can be ascribed to a spherical distribution of the isothermal King model.
The coexistence of torus-like and spherical distributions of the plasma suggests a transition from radial accretion to rotational accretion around the nucleus.
\end{abstract}

\section{Introduction}

The giant elliptical galaxy NGC 1052, which is also classified as a low-ionization nuclear emission-line region (LINER) galaxy, holds an AGN in the center (\cite{Ho1997}).
If we assume $H_0 = 75$ km s$^{-1}$ Mpc$^{-1}$ and $q_0 = 0.5$, the redshift, $z = 0.0049$ (\cite{Knapp1978}), corresponds to a distance of 20 Mpc, and 1 milliarcsec (mas) corresponds to 0.095 pc.
At radio wavelengths, double-sided jets elongated by several pc in P.A.$\sim 65^{\circ}$ are seen with an emission gap in between (\cite{Claussen1998}; \cite{Kellermann1998}), though it is unclear where the nucleus is located.
Water masers are found along the jet, which are considered to be excited by shocks into circumnuclear molecular clouds, or amplification of the radio continuum emission of the jet by foreground molecular clouds (\cite{Claussen1998}).
This object shows a convex radio spectrum peaked at 10 GHz, to be classified as a GHz-Peaked Spectrum (GPS) source (\cite{ODea1998}; \cite{deVries1997}).
A steep spectrum with $\alpha_0 = -1.33$ ($S_{\nu} \propto \nu^{+\alpha_0}$) at frequencies above the peak (\cite{deVries1997}) and a high brightness temperature of $\sim 10^9$ K (see subsection 3.1) indicate synchrotron radiation.

Two possible hypotheses are proposed for the rising spectrum below the peak frequency in GPS sources: synchrotron self-absorption (SSA) or free--free absorption (FFA).
Both can generate convex spectra: $S_{\nu} \propto \nu^{2.5} \left[ 1 - \exp (-\tau_{\rm s} \nu^{\alpha_0 - 2.5}) \right]$ for SSA, and $S_{\nu} \propto \nu^{\alpha_0} \exp (-\tau_{\rm f} \nu^{-2.1})$ for FFA, where $\nu$ is the frequency in GHz, $\tau_{\rm s}$ and $\tau_{\rm f}$ are the SSA and FFA coefficients, respectively, and $\alpha_0$ is the intrinsic spectral index.
In general, the spectral shape will be a product of both.
FFA arises from an external absorber in front of the synchrotron emitter, while SSA is an internal process of the synchrotron emission via an interaction between the relativistic plasma and the magnetic field.
Because these spectral shapes resemble each other, precise spectral measurements are necessary to detect any small difference.
FFA can generate a steeper cutoff than the maximum attainable spectral index, $\alpha$, of $2.5$ by SSA.
Thus, a component with $\alpha > 2.5$ could be evidence of FFA.
Another distinction is related to the difference of the peak frequencies in twin jets.
In the case of SSA, the peak frequency in the receding jet should be lower than that in the approaching jet due to the Doppler shift, if we assume an intrinsic symmetry.
On the contrary, a longer path length towards the receding jet causes a deeper FFA opacity, and thus a higher peak frequency.
A sufficiently high spatial resolution is crucial to avoid spectral blending, as demonstrated by \citet{Kameno2000}.

Because the FFA coefficient is related to the electron density, $n_{\rm e}$ (cm$^{-3}$), and the electron temperature, $T_{\rm e}$ (K), by
\begin{eqnarray}
\tau_{\rm f} = 0.46 \int_{\rm LOS} n_{\rm e}^2 T_{\rm e}^{-3/2} dL, \label{eqn1:ffaopacity}
\end{eqnarray}
where $\int_{\rm LOS} dL$ is the integration via the line of sight in pc, it can be a probe for cold dense plasma (\cite{Pacholczyk1970}).

There exists increasing evidence for FFA towards active galaxies.
One remarkable milestone is the discovery of a counter-jet in 3C 84 at the center of the nearby Seyfert/cD galaxy NGC 1275 (\cite{Vermeulen1994}; \cite{Walker1994}).
The counter-jet showed a strongly inverted spectrum, unlike the approaching jet with a steep spectrum, whose feature can be explained by FFA due to ambient cold dense plasma (see the Appendix).
This idea was confirmed by following VLBA observations by \citet{Walker2000} and VSOP observations by \citet{Asada2000}.

FFA towards GPS sources was predicted by \citet{Bicknell1997}.
They suggested that photoionization of the interstellar medium (ISM) by radiative shocks produces ionized gas, which has a significant emission measure and a correspondingly high FFA opacity.
The discovery of FFA towards GPS sources was made by a space-VLBI observation towards a GPS galaxy, OQ 208, which showed convex spectra towards two radio lobes separated by 10 pc.
The difference in the path length along the line of sight between two lobes causes an asymmetry in the FFA opacity, with a larger opacity towards the far-side lobe.
The FFA opacity implied that cold ($T_{\rm e} \sim 10^5$ K) dense ($n_{\rm e} \sim 10^5$ cm$^{-3}$) plasma associates with the central tens of pc (\cite{Kameno2000}). 
This result showed that multi-frequency VLBI observations are crucial to clarify the FFA, which can be a probe of matter in the vicinity of AGN.
The discovery of the FFA absorber was followed by observations towards the radio galaxies NGC 4261 (\cite{Jones2000}) and 0108+388 (\cite{Marr2000}).

We emphasize two advantages for the FFA as a probe of accreting matter, when the absorber covers the background bright core and jets.
First, since the absorber must be at the near side of the jet, it allows us to figure a 3-dimensional geometry of the system.
Second, the opacity provides physical properties of the absorber, such as the electron density, $n_{\rm e}$, and the temperature, $T_{\rm e}$.
NGC 1052, the nearest GPS source showing the double-sided jets, is one of the best target sources.

In this paper, we report on simultaneous trichromatic Very Long Baseline Array (VLBA) observations towards NGC 1052 at 2.3, 8.4, and 15.4 GHz, which produce an FFA opacity map.
Our aim is to investigate the geometry and physical properties of the FFA absorber.

\begin{figure*}
  \begin{center}
    \FigureFile(172.8mm,204.8mm){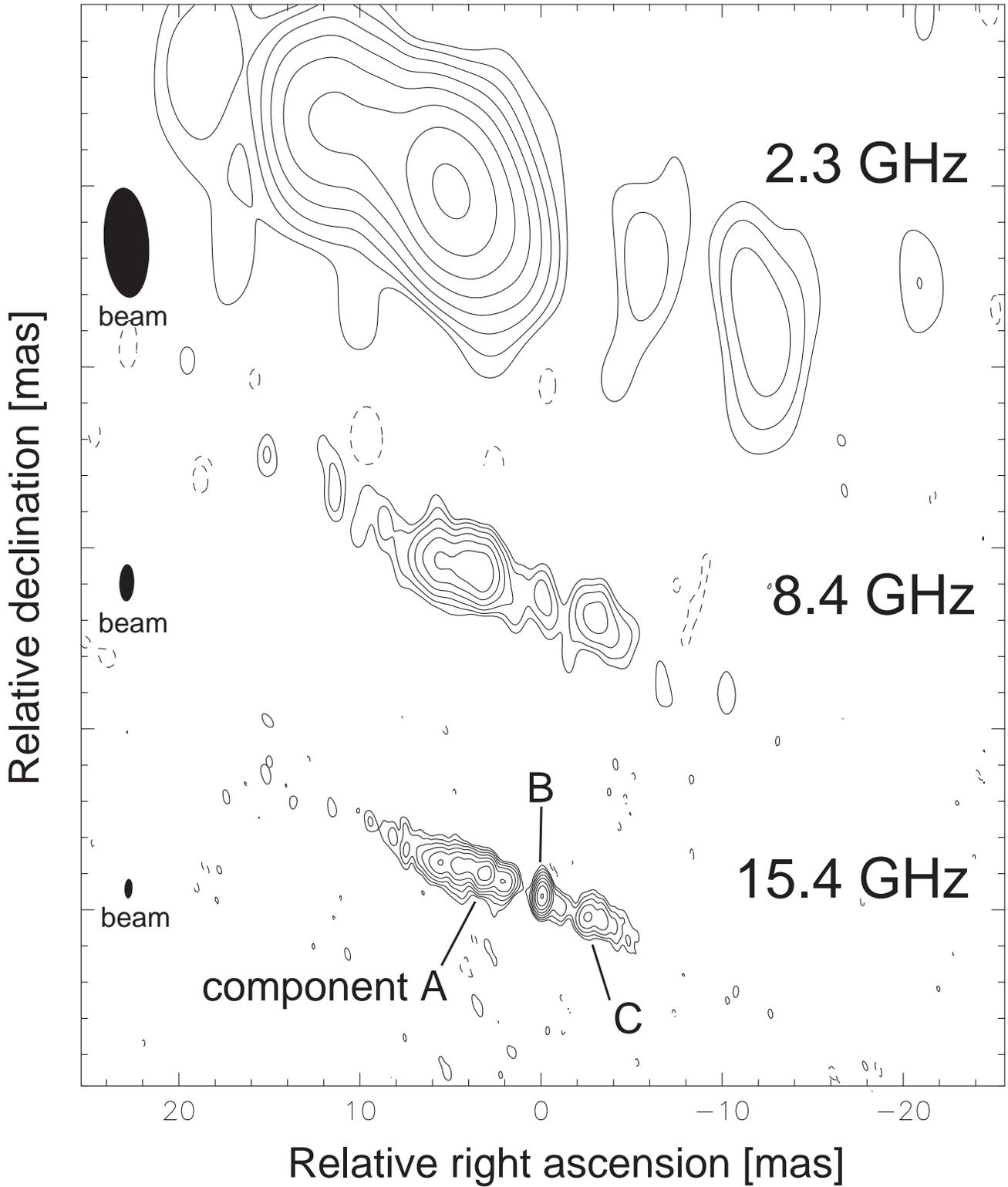}
  \end{center}
  \caption{Images of NGC 1052 at 2.3, 8.4, and 15.4 GHz, taken using VLBA on 1998 December 15.
The synthesized beam sizes (FWHM) are $6.06 \times 2.45$ mas in P.A.$=3^{\circ}.92$, $2.01 \times 0.78$ mas in P.A.$=-1^{\circ}.91$, and $1.03 \times 0.40$ mas in P.A.$=-2^{\circ}.53$, respectively, as shown in the left of each image.
The contours start at the $\pm 3 \sigma$ level, increasing by a factor of 2, where $\sigma = 1.246$, $2.277$, and $0.823$ mJy beam$^{-1}$ for 2.3, 8.4, and 15.4 GHz, respectively.
}\label{fig:Threeimages}
\end{figure*}

\section{Observations and Data Reduction}

Using the VLBA of the National Radio Astronomy Observatory (NRAO), we observed NGC 1052 at three frequencies (2.3, 8.4, and 15.4 GHz) simultaneously on 1998 December 15.
The subreflector switched between the dual-frequency 2.3/8.4 GHz reflector system and the 15.4 GHz feed horn with a 22-min cycle.
We used 4 channels of 8 MHz bandwidth at 15.4 GHz, and allocated 2 channels at both 2.3 and 8.4 GHz.
The correlation process was accomplished by the VLBA correlator.
We applied fringe-fitting, data flagging, and a priori amplitude calibration in the NRAO AIPS (Astronomical Image Processing System).
Imaging and self-calibration processes were carried out using Difmap (\cite{Shepherd1997}).
We took uniform weighting with scaling by errors raised to the power $-1$.
The synthesized beam sizes (FWHM) were $6.06 \times 2.45$ mas in P.A.$=3^{\circ}.92$, $2.01 \times 0.78$ mas in P.A.$=-1^{\circ}.91$, and $1.03 \times 0.40$ mas in P.A.$=-2^{\circ}.53$, for 2.3, 8.4, and 15.4 GHz, respectively.

While relative gain errors among the antennas are corrected though amplitude self-calibration processes, further flux calibration is necessary to obtain certain spectra across the observing frequencies. 
For the purpose of the absolute flux calibration, we also imaged four calibrators: the BL Lacertae, DA 193, 3C 279, and OT 081.
Based on a comparison between the total flux measurements by the University of Michigan Radio Astronomy Observatory (UMRAO) and the NRAO Green Bank interferometer, and a summation of the CLEANed flux densities, we applied flux scaling for the final results.  
At 15.4 GHz, for instance, the total CLEANed flux densities before an absolute correction were 3.276, 5.012, 25.769, and 4.172 Jy for BL Lac, DA 193, 3C 279, and OT 081, respectively.
The UMRAO database provides the total flux densities at 14.5 GHz, $3.458 \pm 0.015$, $4.763 \pm 0.035$, $27.69 \pm 0.18$, and $4.380 \pm 0.030$ Jy for each, as averaged over 2 months centered at our observation.
We then calculated a correction factor ($= \mbox{total CLEANed flux density} / \mbox{UMRAO flux density}$) of $0.959 \pm 0.018$, so that we scaled our image by the factor and obtained an accuracy of the flux scale of 1.8\%.
The accuracy was derived from a standard deviation of the correction factors by the four calibrators. 
In the same manner, we scaled images at 2.3 and 8.4 GHz by correction factors of 0.84 and 0.79, respectively.
The estimated amplitude accuracies are 4.0, 7.2, and 1.8\% at 2.3, 8.4, and 15.4 GHz, respectively.
The images are shown in figure \ref{fig:Threeimages}.

\begin{figure*}
  \begin{center}
    \FigureFile(165.6mm,88.5mm){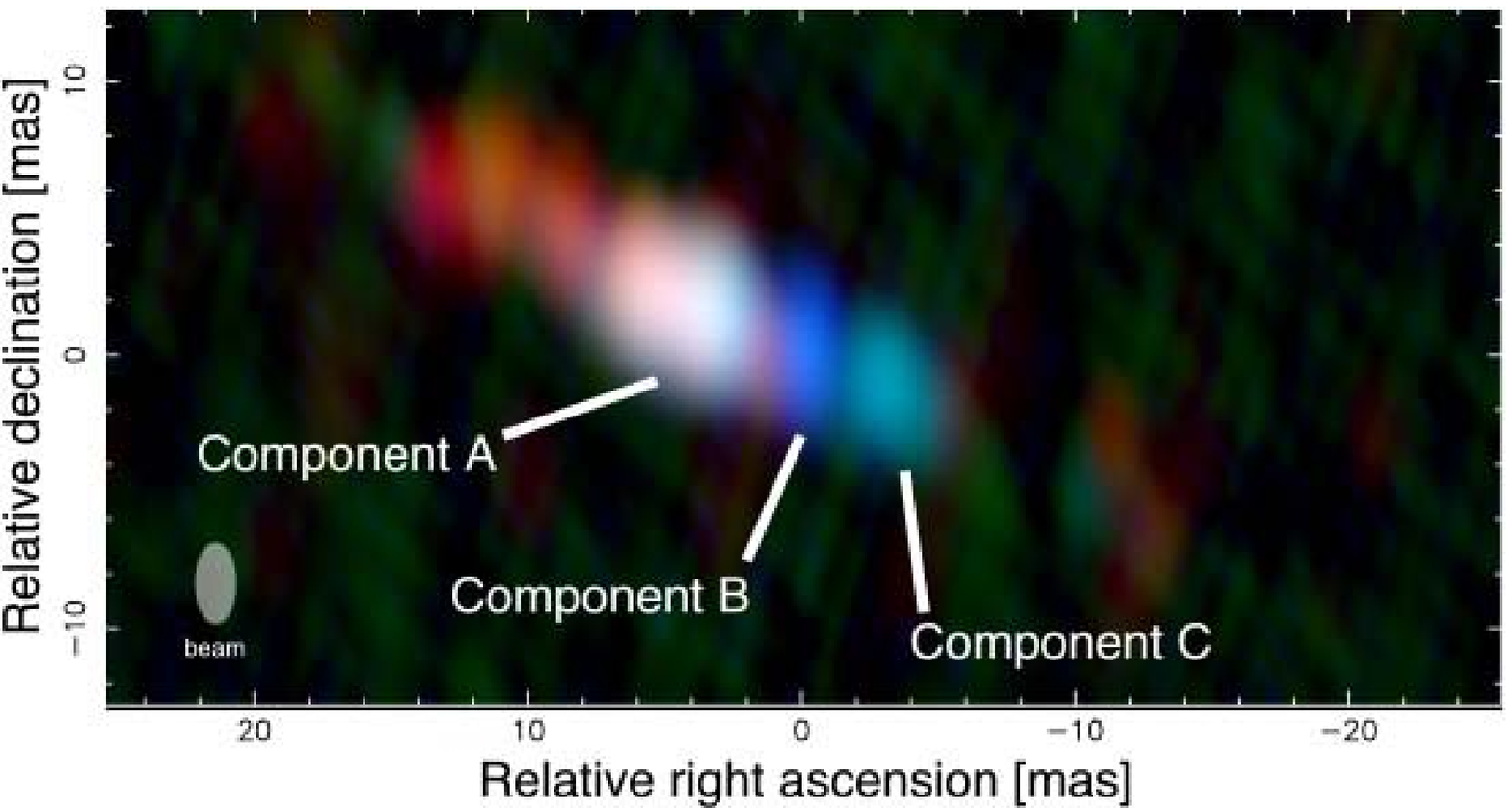}
  \end{center}
  \caption{Trichromatic color image of NGC 1052 synthesized after restoring the images at 2.3, 8.4, and 15.4 GHz with a $3 \times 1.5$ mas beam, and register by referencing components A and C.
The red, green, and blue color channels are allocated for the images at 2.3, 8.4, and 15.4 GHz, respectively.
The origin of all images are referred by the position of component B, which is considered to be the nucleus.
The blue part covering the nucleus and $\sim 5$ mas to the west indicates a steeply rising spectrum with $\alpha > 0$ due to large FFA opacities.
The red color at both ends of the jets shows $\alpha < 0$, suggesting optically-thin synchrotron emission.}
\label{fig:Trichromat}
\end{figure*}

\begin{longtable}{lrccccc}
  \caption{Parameters of the Gaussian components.}\label{tab:components}
  \hline\hline
\multicolumn{1}{c}{Comp.}&
\multicolumn{1}{c}{Frequency} &
\multicolumn{1}{c}{Flux Density} &
\multicolumn{1}{c}{Relative R. A.$^*$} &
\multicolumn{1}{c}{Relative Dec.$^*$} &
\multicolumn{1}{c}{$\phi_{\rm maj}^{\dag}$} &
\multicolumn{1}{c}{$\phi_{\rm min}^{\dag}$} \\ 
{~~~~~~~~~~~~~~~~~}& 
\multicolumn{1}{c}{~~(GHz)~~} &
\multicolumn{1}{c}{~~(mJy)~~} &
\multicolumn{1}{c}{~~(mas)~~} &
\multicolumn{1}{c}{~~(mas)~~} &
\multicolumn{1}{c}{~~(mas)~~} &
\multicolumn{1}{c}{~~(mas)~~} \\
\hline
\endfirsthead
\endhead
  \hline
\endfoot
  \hline
\multicolumn{7}{l}{$^{*}$Relative position with respect to component A.} \\
\multicolumn{7}{l}{$^{\dag}$Full width at half maximum (FWHM) of the major and minor axes of the Gaussian component.} \\
\endlastfoot
A\dotfill & 2.274 & $969\pm39$   & $0$     & $0$     & $4.2$ & $1.2$ \\
          & 8.424 & $1800\pm130$ & $0$     & $0$     & $3.4$ & $0.9$ \\
          & 15.364& $1310\pm24$  & $0$     & $0$     & $3.4$ & $0.9$ \\
B\dotfill & 2.274 & $<5.6$       & $\cdots$ & $\cdots$ & $\cdots$ & $\cdots$ \\
          & 8.424 & $77.0\pm5.7$ & $-4.59 \pm 0.2$ & $-2.03 \pm 0.3$ & $<1.3$ & $<0.7$ \\
          & 15.364& $488.7\pm8.8$& $-3.59 \pm 0.2$ & $-1.36 \pm 0.3$ & $<0.5$ & $<0.4$ \\
C\dotfill & 2.274 & $12.6\pm2.2$ & $-8.85 \pm 0.7$ & $-3.44 \pm 1.8$ & $2.1$ & $1.4$ \\
          & 8.424 &$247.2\pm18.0$& $-7.16 \pm 0.2$ & $-2.91 \pm 0.3$ & $1.6$ & $0.6$ \\
          & 15.364& $189.3\pm3.8$& $-6.29 \pm 0.2$ & $-2.53 \pm 0.3$ & $1.6$ & $0.6$ \\ 
\end{longtable}

\section{Discussion}

\subsection{Component Identification and Spectrum}
For the purpose of registration, we picked up three distinct components (A, B, and C ;see figure \ref{fig:Threeimages}).
Using the task `IMFIT' in AIPS, we measured the relative positions, sizes and flux densities of these components, as listed in table \ref{tab:components}.
In advance of the IMFIT process, we restored the CLEAN images with a Gaussian beam of $3 \times 1.5$ mas in P.A.$ = 0^{\circ}$ to match the resolutions.
Double-sided jet components, A and C, are identified with similar separations of 9.5, 7.7, and 6.8 mas at 2.3, 8.4, and 15.4 GHz, respectively.
An unresolved component, B, is clearly seen with a significant flux density of $489 \pm 9$ mJy at 15.4 GHz, but becomes fainter $77 \pm 6$ mJy at 8.4 GHz and too faint ($< 5.6$ mJy) to be identified at 2.3 GHz.

These components allow us to register three images to minimize the total difference of their positions.
At any given frequency, we measured the positions $(\xi_k^{\nu}, \eta_k^{\nu})$ of these three components relative to a tentative origin of each image, where $k$ and $\nu$ stand for an index of the components (A, B, and C) and a frequency, respectively.
We then derived the relative offsets $(\delta \xi^{\nu}, \delta \eta^{\nu})$ between the images to minimize the positional residuals, $\chi_{\nu}^2$, defined as
\begin{eqnarray}
\chi_{\nu}^2 = \sum_{k} \left[ \frac{ (\xi_k^{\nu} - \delta \xi^{\nu} - \xi_k^{\nu_0})^2}{\sigma^2(\xi_k^{\nu}) + \sigma^2(\xi_k^{\nu_0})} + \frac{ (\eta_k^{\nu} - \delta \eta^{\nu} - \eta_k^{\nu_0})^2}{\sigma^2(\eta_k^{\nu}) + \sigma^2(\eta_k^{\nu_0})}  \right],
\end{eqnarray}
where $\sigma(\xi_k^{\nu})$ and $\sigma(\eta_k^{\nu})$ are standard positional errors of component $k$ at frequency $\nu$.
The reference frequency, $\nu_0$, was put at 8.4 GHz.
The r.m.s. errors of the registration in $(\xi, \nu)$ were $(0.25, 0.09)$ mas between 15.4 and 8.4 GHz, and $(0.20, 0.25)$ mas between 8.4 and 2.3 GHz, respectively.
The response to the registration errors is discussed in subsection 3.4.

Using image registration, we synthesized a trichromatic image (figure \ref{fig:Trichromat}) allocating red, green, blue channels for 2.3, 8.4, and 15.4 GHz, respectively, using the beam-matched images.
The blue color covering component B and western $\sim 5$ mas indicates a steeply rising spectrum ($\alpha > 0$), while the red color at both ends of the elongation shows $\alpha < 0$.

\begin{figure}
  \begin{center}
    \FigureFile(80.25mm,126.0mm){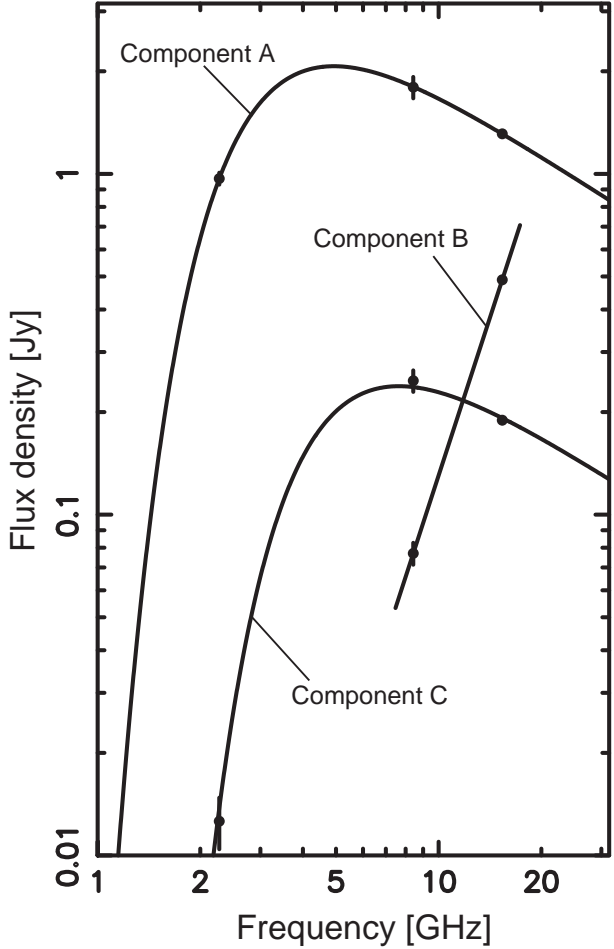}
  \end{center}
  \caption{Spectra of components A, B, and C.
The flux densities were measured by a Gaussian model fit using the task `IMFIT' in AIPS, as listed in table \ref{tab:components}.
The errors includes both formal errors and amplitude calibration errors.
The solid lines for components A and C show the best-fit FFA model, $S_{\nu} = S_0 \nu^{\alpha_0} \exp (-\tau_{\rm f} \nu^{-2.1})$, where $S_{\nu}$ is the observed flux density in Jy, $S_0$ is the unabsorbed flux density at 1 GHz, $\alpha_0$ is the intrinsic spectral index, $\tau_{\rm f}$ is the FFA opacity coefficient, and $\nu$ is the frequency in GHz.
}
\label{fig:Spectrum}
\end{figure}

\begin{figure}
  \begin{center}
    \FigureFile(73.75mm,141.5mm){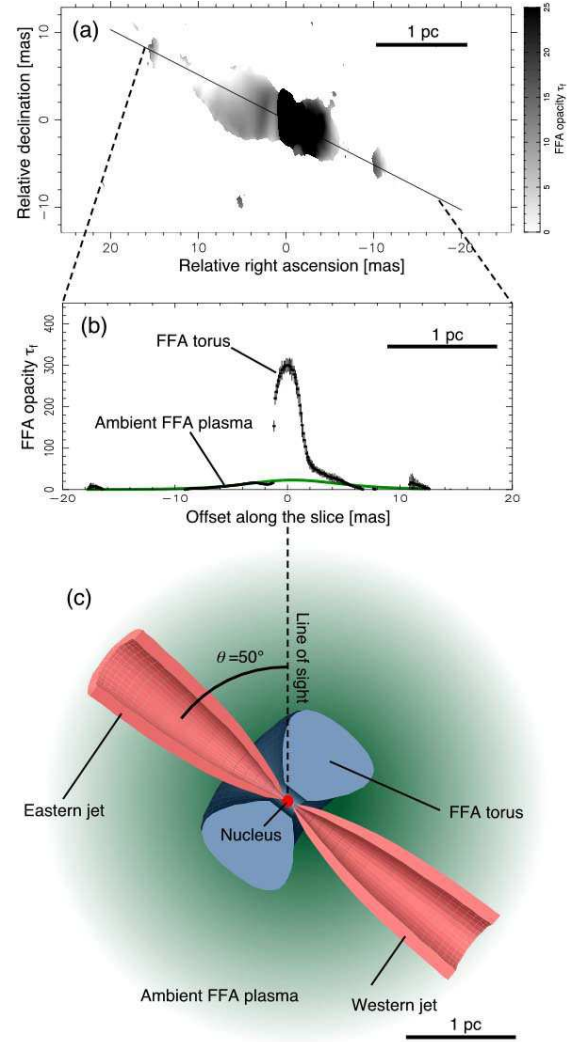}
  \end{center}
  \caption{Distribution of the FFA opacity and a schematic model of a torus surrounding NGC 1052.
(a): The distribution of the FFA opacity, $\tau_{\rm f}$, is drawn in grey scale.
The opacity was calculated by the least-squares fit for the FFA model, $S_{\nu} \propto \nu^{\alpha_0} \exp (-\tau_{\rm f} \nu^{-2.1})$.
It was obtained by the trichromatic image (figure \ref{fig:Trichromat}), which was restored with a $3 \times 1.5$ mas beam.
The fit was applied at pixels of $0.1 \times 0.1$ mas with brightness thresholds of $5 \sigma$ at 8.4 and 15.4 GHz.
Pixels below $5 \sigma$ at 2.3 GHz (and above $5 \sigma$ at 8.4 and 15.4 GHz) are valid for the fit; we put the $5 \sigma$ value as an upper limit.
We clipped out pixels where the FFA fit did not result in a positive $\tau_{\rm f}$.
(b): The opacity profile along the jet in P.A.$=65^{\circ}$.
It peaks at $\tau_{\rm f} \sim 300$ in the nucleus.
It tails over $\sim 0.7$ pc along the western (receding) jet, and rapidly falls at 0.1 pc in the eastern (approaching) jet.
Out of this region, an extended profile of an ambient FFA absorber can be seen.
The green curve shows the best-fit model of the FFA opacity by the ambient plasma with the isothermal King model distribution described as equation (\ref{eqn:KingNe}).
(c): A schematic diagram of NGC 1052.
The nucleus and the double-sided jets are visible at 15.4 GHz, with the viewing angle of $50^{\circ}$.
The ambient plasma (green) distributes around the nucleus supported by its thermal pressure, which contributes an extended FFA components.
The FFA torus is presumed to lie perpendicular to the double-sided jets.
It covers $\sim 0.7$ pc of the receding jet and $\sim 0.1$ pc of the jet, to generate the asymmetry seen in the opacity distribution.
}
\label{fig:schematic}
\end{figure}

Figure \ref{fig:Spectrum} shows the spectra of the components.
Both A and C show convex spectra, which can be fitted by optically-thin synchrotron emission obscured by the cold plasma via thermal FFA.
The spectral index, $\alpha_0 = -1.33$ (\cite{deVries1997}), was too steep to fit, instead, we obtained $\alpha_0 = -0.65$ from the best fit.
A and C mark the spectral peaks at 4.9 and 7.6 GHz, respectively, while B must have a peak above $15$ GHz.
Although the spectra of A and C seem to be similar, the difference of the peak frequencies is significant.
The colors of these components in the trichromatic image (figure \ref{fig:Trichromat}) emphasize the difference.
The spectral indices between 2.3 and 8.4 GHz are $\alpha_{2.3}^{8.4} = 0.5 \pm 0.1$ and $2.3 \pm 0.1$ for A and C, and $\alpha_{\rm 2.3}^{\rm 8.4} > 2.0$ for B.
Component B shows a large spectral index of $\alpha_{8.4}^{15.4} = 3.1 \pm 0.1$ between 8.4 and 15.4 GHz.
It exceeds the theoretical limit for SSA of 2.5 with a uniform magnetic field (\cite{Pacholczyk1970}).
A and C are resolved at 15.4 GHz with brightness temperatures of $T_{\rm b} = (2.32 \pm 0.04) \times 10^9$ K and $(0.97 \pm 0.02) \times 10^9$ K, respectively, while component B is unresolved and shows the highest $T_{\rm b} > 1.48 \times 10^{10}$ K.
Thus, we identify component B as the nucleus.

\subsection{Viewing Angle of the Jet Axis}
To clarify the geometry of the jet, we analyze the Doppler beaming effect (\cite{Blandford1997}; \cite{Pearson1987}).
Although both FFA and the Doppler beaming result in an apparent asymmetry of brightness between approaching and receding jets, the former is negligible at frequencies beyond the peak.
If we assume the intrinsic symmetry of the double-sided jets, the brightness ratio, $R$, is related to the intrinsic jet velocity, $\beta$ (in the unit of the speed of light), and the viewing angle, $\theta$, by
\begin{eqnarray}
R = \frac{T_{\rm b}^{+}}{T_{\rm b}^{-}} = \left( \frac{1 + \beta \cos \theta}{1 - \beta \cos \theta} \right) ^{2-\alpha_0}.
\end{eqnarray}
Here, $T_{\rm b}^{+}$ and $T_{\rm b}^{-}$ are the brightness temperature of the approaching and receding jets, respectively.
Substituting $T_{\rm b}^{+} = (2.32 \pm 0.04) \times 10^9$ K and $T_{\rm b}^{-} = (0.97 \pm 0.02) \times 10^9$ K of components A and C, respectively, we have $R=2.4 \pm 0.1$ at 15 GHz, and $\beta \cos \theta = 0.16 \pm 0.01$ for $\alpha_0 = -0.65$.
The apparent proper motion of the jet gives another constraint.
The relative proper motion of $1.3$ mas yr$^{-1}$ between the approaching and receding jets (\cite{Kellermann1999}) corresponds to $\beta_{\rm app} = \beta_{\rm app}^{+} + \beta_{\rm app}^{-} = 0.40$ .
Here, the positive and negative signs are for the approaching and receding jets, respectively.
Using the formula of the apparent velocity (\cite{Blandford1997}),
\begin{eqnarray}
\beta_{\rm app}^{\pm} = \frac{\beta \sin \theta}{1 \mp \beta \cos \theta},
\end{eqnarray}
we have $\beta \sin \theta = \beta_{\rm app} (1 - \beta^2 \cos^2 \theta) / 2$.
Finally, we obtain $\beta=0.25$ and $\theta = 50^{\circ}$.
Although this is a rough estimation, it is obvious that the eastern and western jets are approaching and receding, respectively.
A proper-motion analysis based on the VLBA monitoring program at 15 GHz (\cite{Kellermann1998}; \cite{Kellermann1999}) will confirm these results and evaluate the accuracy.

\begin{figure}
  \begin{center}
    \FigureFile(80.25mm,51.9mm){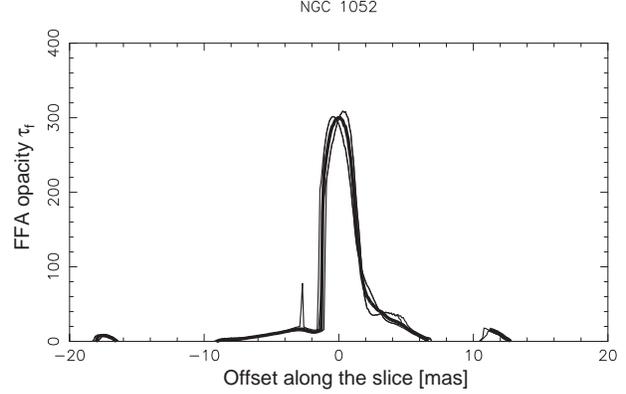}
  \end{center}
  \caption{Tolerance of the FFA opacity profile along the jet against the position errors in registration.
The thick line is the same as in figure \ref{fig:schematic}b, which shows the opacity profile with nominal registration. 
The thin lines represents the results of four test cases, with shifts at 15.4 and 2.3 GHz, $(\delta \xi^{\rm 15.4}, \delta \xi^{\rm 2.3})$ of $(+0.2,\, +0.2)$, $(+0.2,\, -0.2)$, $(-0.2,\, +0.2)$, $(-0.2,\, -0.2)$ mas.
These tentative profiles align with that of the nominal registration, representing the peak towards the nucleus, the tail to the west receding jet, and the break at 1 mas to the east approaching jet. 
A spike of $\tau_{\rm f} = 78$ at $-2.7$ mas from the nucleus can be seen only in the case of $(\delta \xi^{\rm 15.4}, \delta \xi^{\rm 2.3}) = (-0.2,\, +0.2)$ mas, which reflects on the tolerance of the registration when opposite registration errors for 2.3 and 15.4 GHz.
}
\label{fig:tolerance}
\end{figure}

\subsection{The Absorption Mechanisms}
We discuss the origin of the low-frequency cutoff in the spectra: SSA or FFA.
In the SSA process, the opacity is related to the magnetic field, $B$, and brightness temperature, $T_{\rm b}$, at the peak frequency.
If we assume equipartition between the magnetic field and the electrons, the peak frequency, $\nu_{\rm m}$, in GHz is given by \citet{Kellermann1981},
\begin{eqnarray}
\nu_{\rm m} \sim 8 B^{1/5} S_{\rm m}^{2/5} \phi^{-4/5} (1+z)^{1/5}.\label{eqn:ssapeak}
\end{eqnarray}
Here, the magnetic field, $B$, is in G, the peak flux density, $S_{\rm m}$, is in Jy, the component size, $\phi$, is in mas.
Equation (\ref{eqn:ssapeak}) is replaced by
\begin{eqnarray}
B = 4.57 \times 10^{19} \nu_{\rm m} T_{\rm b}^{-2} (1+z),
\end{eqnarray}
using the brightness temperature of the Gaussian component, $T_{\rm b}$, in K,
\begin{eqnarray}
T_{\rm b} = 1.224 \times 10^{12} S \nu^{-2} \phi^{-2} (1+z).
\end{eqnarray}
For the approaching jet component, A, the peak flux density, $S_{\rm m} = 2.0$ Jy, and the size of $\phi = \sqrt{\phi_{\rm maj} \phi_{\rm min}} = 2$ mas at the peak frequency, $\nu_{\rm m} = 4.9$ GHz, give $T_{\rm b} = 2.6 \times 10^{10}$ K.
In the co-moving frame, the peak frequency is modified as $\nu_{\rm m}^{*} = 4.3$ GHz by the Doppler correction,
\begin{eqnarray}
\nu_{\rm m}^{*} = \nu_{\rm m} (1+z) \frac{1 - \beta \cos \theta}{\sqrt{1 - \beta^2}}.
\end{eqnarray}
The brightness temperature in the co-moving frame, $T_{\rm b}^{*}$, is also corrected by
\begin{eqnarray}
T_{\rm b}^{*} = T_{\rm b} \left( \frac{1 - \beta \cos \theta}{\sqrt{1 - \beta^2}} \right)^{2 - \alpha_0},
\end{eqnarray}
and $T_{\rm b}^{*} = 1.5 \times 10^{10}$ K is obtained.
We thus have $B = 0.9$ G.

For the receding jet component, C, we have $\nu_{\rm m} = 7.6$ GHz, $S_{\rm m} = 0.25$ Jy, $\phi = 1.0$ mas, and $T_{\rm b} = 5.3 \times 10^{9}$ K.
These values should be corrected as
\begin{eqnarray}
\nu_{\rm m}^{*} &=& \nu_{\rm m} (1+z) \frac{1 + \beta \cos \theta}{\sqrt{1 - \beta^2}}, \\
T_{\rm b}^{*} &=& T_{\rm b} \left( \frac{1 + \beta \cos \theta}{\sqrt{1 - \beta^2}} \right)^{2 - \alpha_0}.
\end{eqnarray}
Thus, we have $\nu_{\rm m}^{*} = 9.2$ GHz and $T_{\rm b}^{*} = 8.6 \times 10^{9}$ K; therefore $B=5.7$ G.
Such strong magnetic fields are unrealistic and impossible to sustain synchrotron emission for years against synchrotron loss.
If we assume that components A and C have been emitting for 20 yr (= separation / apparent speed) without re-acceleration, $B < 55$ mG is required to sustain synchrotron emission with a break frequency of $\nu_{\rm T} > 15.4$ GHz.
In this case, $\nu_{\rm m} < 260$ MHz.
Thus, the synchrotron emission is likely to be optically thin, in terms of SSA, towards the jets throughout our observing frequencies.

Since SSA is an internal process in the jet, a difference in the peak frequencies at the rest frame requires an intrinsic asymmetry of the magnetic fields between the approaching and receding jets.
On the contrary, FFA naturally generates a higher peak frequency towards the receding jet than towards the approaching one.
In the case of a spherically symmetric distribution of the external absorber, the path length along the line of sight will be longer towards the receding jet than towards the approaching one (see figure \ref{fig:ambientmodel}).
Furthermore, if the absorber forms a disk or a torus perpendicular to the jet, it covers the receding jet, but does not cover the approaching jet.
Such a geometry is discussed in the next section.

Although $T_{\rm b}$ and $\nu_{\rm m}$ are uncertain for component B, the magnetic field can be roughly estimated.
Since the total flux density of $1.02 \pm 0.12$ Jy at 31.4 GHz (\cite{Geldzahler1981}) is less than a natural extension of the rising spectrum of component B, the peak frequency, $\nu_{\rm m}$, should be $15.4$ GHz $< \nu_{\rm m} < 31.4$ GHz.
The intrinsic brightness temperature should be less than the Compton limit, $T_{\rm b} < 10^{12}$ K.
Therefore, $7 \times 10^{-4}$ G $< B < 7$ G.
Although the nucleus component, B, could be optically thick by SSA, its spectral index of $\alpha_{8.4}^{15.4} = 3.1 \pm 0.1$ requires a significant contribution of FFA.
To keep simplicity, we later analyze the spectra assuming that the synchrotron emission is optically thin in terms of SSA, rather than a mixture of SSA and FFA.
Our FFA analysis would be modified if the SSA opacity were revealed by future observations.

In some GPS sources, the SSA model outputs an acceptable strength of the magnetic fields.
\citet{Mutel1985} calculated magnetic fields of $\sim 10^{-4} - 10^{-3}$ G from SSA for three sources: 1518+047, CTD 93, and 2050+364. 
\citet{Readhead1996} estimated magnetic fields of $3 \times 10^{-2} h^{2/7}$ G and $4 \times 10^{-2} h^{2/7}$ for the bright jet components, $B_1$ and $B_2$, of a GPS source 2352+495, based on the brightness temperatures of $T_{\rm b} = 2 \times 10^{10}$ K and $2 \times 10^{11}$ K, and a peak frequency of $\nu_{\rm m} \sim 1.66$ GHz, with a Hubble constant, $H_0$, of $100 h$ km s$^{-1}$ Mpc$^{-1}$.
\citet{Shaffer1999} derived $B=10^{-4}$ G for the lobes of a GPS source, CTD 93, based on $T_{\rm b} = 1.3 \times 10^{11}$ K and $0.7 \times 10^{11}$ K for the northern and southern lobes, respectively, and $\nu_{\rm m} \sim 1$ GHz.
A higher $T_{\rm b}$ and a lower $\nu_{\rm m}$ than those of NGC 1052 result in the much lower $B$ required from the SSA model for these sources. 

\citet{Shaffer1999} claimed that the observed dependence of the apparent component separation of CTD 93 on wavelength is additional evidence for SSA.
NGC 1052 also shows such a frequency dependence of the apparent component separation (see table \ref{tab:components}); nevertheless, we mention that the central condensation of the FFA opacities can explain it.

\begin{figure}
  \begin{center}
    \FigureFile(81.5mm,65.5mm){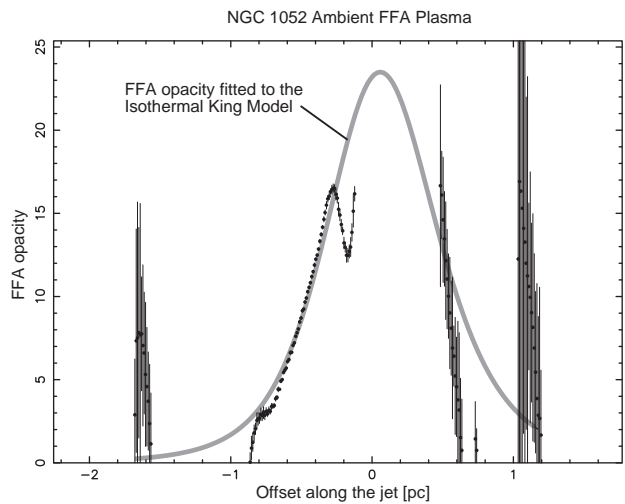}
  \end{center}
  \caption{FFA opacities of NGC 1052 out of the torus.
The horizontal axis indicate the offset along the jet in P.A.$=65^{\circ}$ in pc, where 1 mas corresponds to 0.1 pc.
The negative and positive offsets stand for the eastern (approaching) and western (receding) jets, respectively.
The vertical axis shows the FFA coefficient, $\tau_{\rm f}$, derived from the FFA model fit.
The opacities in the torus, $ -1 < x < 4$ mas (or -0.1 to 0.4 pc), were removed from the plot.
The solid grey line shows the opacity calculated from the isothermal King model, which is calculated in the Appendix and shown in figure \ref{fig:Kingmodel}.
Here, we take the viewing angle $\theta = 50^{\circ}$.
The best fit gives $n_0 = (1.00 \pm 0.03) \times 10^4 ( T_{\rm e} / 10^4 {\rm K} )^{3/4} $ cm$^{-3}$ and $r_{\rm c} = 0.84 \pm 0.02$ pc.
}
\label{fig:ambient}
\end{figure}

\subsection{Free--Free Absorption Opacity Distribution}
By applying a least-squares fit for the FFA spectrum at every image pixel, we obtain the distribution of the FFA opacity, $\tau_{\rm f}$, shown in figure \ref{fig:schematic}a.
The fit is applied for restored images at all three frequencies with a $3.0 \times 1.5$ mas beam in P.A.$=0^{\circ}$.
The opacity map is clipped by the thresholds for the brightness of $5 \sigma$ at 8.4 and 15.4 GHz.
Pixels below $5 \sigma$ at 2.3 GHz (and above $5 \sigma$ at 8.4 and 15.4 GHz) are valid for the fit; we put the $5 \sigma$ value as an upper limit.
We reject pixels where the FFA fit does not result in a positive $\tau_{\rm f}$.
The intrinsic spectral index, $\alpha_0$, is fixed to be $-0.65$ at every pixel, since three frequencies are insufficient to allow any freedom in $\alpha_0$.
It should be mentioned that the FFA opacity would be overestimated if $\alpha_0$ were larger in the nucleus.

Figure \ref{fig:schematic}b shows a slice of the distribution of the FFA opacities along the jet in P.A.$=65^{\circ}$.
The error bars indicate formal errors of the least squares fit for the FFA spectrum.
The errors of registration between images at different frequencies, derived in subsection 3.1, can contribute to additional uncertainties of the FFA opacities.
To evaluate the uncertainties responding to the registration errors, we tentatively shifted images at 2.3 and 15.4 GHz relative to the 8.4-GHz image by 0.2 mas, performed a least-squares FFA fit, and then plotted the opacity profiles along the jet.
The results of four test cases, with the shifts at 15.4 and 2.3 GHz, $(\delta \xi^{\rm 15.4}, \delta \xi^{\rm 2.3})$ of $(+0.2,\, +0.2)$, $(+0.2,\, -0.2)$, $(-0.2,\, +0.2)$, $(-0.2,\, -0.2)$ mas, are plotted in figure \ref{fig:tolerance}.
The profiles align with that of the nominal registration, except for a spike of $\tau_{\rm f} = 78$ at $-2.7$ mas from the nucleus in the case of $(\delta \xi^{\rm 15.4}, \delta \xi^{\rm 2.3}) = (-0.2,\, +0.2)$ mas. 
This spurious spike reflects on the tolerance of the registration, where the brightness distribution gaps between components A and B.
However, we have a small joint probability to meet such a bad case with opposite position errors for 2.3 and 15.4 GHz.
We also attempted position shifts in the north--south direction, and confirmed that the opacity profile is insensitive in this direction, probably due to elongation of the restoring beam.

As shown in equation (\ref{eqn1:ffaopacity}), $\tau_{\rm f}$ can be used to investigate the cold dense plasma.
The opacity peaks at $\tau_{\rm f} \sim 300$ towards the nucleus and decreases to both downstream regions.
In the central sub-pc from the nucleus, we found a significant asymmetric distribution along the jet.
The opacity rapidly falls at 0.1 pc in the eastern jet, while it smoothly decreases over 0.7 pc in the western jet.
This feature can be seen in the trichromatic image (figure \ref{fig:Trichromat}), in which blue color indicates an opaque area at low frequency due to FFA.
As stated before, we find that the eastern and western jets are approaching and receding, respectively, in the viewing angle of $\theta \sim 50^{\circ}$.
The dense absorber does not cover the approaching jet, but does cover the receding jet.
Thes implies that the absorber forms a disk or torus perpendicular to the jet, as illustrated in figure \ref{fig:schematic}c.
In this configuration, the receding jet is covered with the torus, but the approaching jet isn't.
A rapid change of the opacity at 0.1 pc in the approaching jet suggests that the torus is geometrically thick with a thickness $H > r \cot \theta$, where $r$ is the radius of the torus.
Since the opacity tails along the receding jet up to $\sim 0.7$ pc from the nucleus, we consider that the radius of the torus is $\sim 0.5$ pc. 
The maximum opacity, $\tau_{\rm f} = 300$, for a  path length of $\sim 0.7$ pc gives a constraint for $n_{\rm e}$ and $T_{\rm e}$ by $n_{\rm e}^2 T_{\rm e}^{-1.5} = 930$.
The ionization condition of $T_{\rm e} > 10^4$ K constrains $n_{\rm e} > 3.1 \times 10^4$ cm$^{-3}$, and thus the column density, $n_{\rm e} L \sim 0.7 \times 10^{23} (T_{\rm e} / 10^4 \: {\rm K} )^{3/4}$ cm$^{-2}$.
This is consistent with an atomic column density of $\sim 10^{23}$ cm$^{-2}$ derived from the ROSAT and ASCA X-ray observations (\cite{Guainazzi1999}).
Such a high column density is found via  H\,{\footnotesize I} absorption towards some Seyfert and radio galaxies.
Although the H\,{\footnotesize I} column densities hold an ambiguity of the spin temperature, $T_{\rm spin}$, those of $(3.9 \pm 0.5) \times 10^{21} (T_{\rm spin}/100 \, {\rm K})$ cm$^{-2}$ towards NGC 4151 (\cite{Mundell1995}), $(2.54 \pm 0.44) \times 10^{19} T_{\rm spin} $ cm$^{-2}$ towards Cygnus A (\cite{Conway1995}), $14.0 \times 10^{21} (T_{\rm spin}/100 \, {\rm K})$ cm$^{-2}$ towards the core of Hydra A (\cite{Taylor1996}), $27.8 \times 10^{23} (T_{\rm spin}/8000 \, {\rm K})$ cm$^{-2}$ in the broad velocity-width component towards the core of PKS 2322$-$123 (\cite{Taylor1999}), and $(19.3 \pm 3.8) \times 10^{21} (T_{\rm spin}/100 \, {\rm K})$ cm$^{-2}$ towards NGC 3079 (\cite{Sawada2000}) are comparable with the electron column densities probed by FFA.
A comparison of distribution of FFA with that of H\,{\footnotesize I} absorption could illustrate the ionizing surface of the torus.

\subsection{Ambient Free--Free Absorbing Plasma}

Besides the asymmetric condensation in the central 0.1 pc, the opacity widely extends in the outer region, $x < -1.5$ mas, in the approaching jet and $x > 10$ mas in the receding one.
Here, $x$ is the offset from the nucleus along the jet in P.A.$=65^{\circ}$.
The opacity profile in this ambient absorber can be explained by the isothermal King model (\cite{King1972}),
\begin{eqnarray}
n_{\rm e} = n_0 \left[ 1 + \left(\frac{r}{r_{\rm c}}\right)^2 \right]^{-1.5}, \label{eqn:KingNe}
\end{eqnarray}
where $n_0$ is the electron density at the centroid, $r$ the radius from the centroid and $r_{\rm c}$ the core radius.
Making a least-squares fit for the opacities in $x < -1$ and $x > 4$ mas, we estimated $n_0 = (1.00 \pm 0.03) \times 10^4 ( T_{\rm e} / 10^4 {\rm K} )^{3/4} $ cm$^{-3}$ and $r_{\rm c} = 0.84 \pm 0.02$ pc.
The best-fit curve is shown as the green line in figures \ref{fig:schematic}b and \ref{fig:ambient}.
See the Appendix for the analytic profile of the opacities towards the slant jets.

It is remarkable that the torus-like plasma distribution coexists with the spherically symmetric distribution.
The former should be supported by the centrifugal force by rotation around the nucleus, while the latter is supported by the thermal pressure.  
We consider that our results witness the transition stage from the radial accretion to rotational accretion.
We note that water masers (\cite{Claussen1998}) distribute between components B and C, where the FFA torus covers the foreground of the receding jet.
Our results support the idea that the masers are amplified radio continuum emission from the receding jet by the foreground disk.
The velocity gradient along the jet axis, faster receding velocity at a position closer to the nucleus, may represent a dynamical infall in the torus.

\section{Conclusion}

We have performed trichromatic VLBA observations towards the nucleus of NGC 1052.
Based on the peak frequency and brightness temperature, we identified the nucleus component in this object.
The kinematics and orientation of the double-sided jets were derived from the Doppler beaming effect.
We found that the spectra of the core and double-sided jets can be better fitted with the free--free absorption process by the cold dense plasma.
The distribution of the opacity is biased towards the receding jet, suggesting a torus-like condensation perpendicular to the jet.
Out of the torus, we also found an ambient plasma distribution which follows the spherical isothermal King model.
Our results indicate the first look at a dynamical transition from thermally-supported into centrifugally-supported in the subparsec-scale region of AGN.

\par
\vspace{1pc}\par
We thank M. Miyoshi for helpful discussions.
The VLBA is operated by the National Radio Astronomy Observatory, which is a facility of the National Science Foundation (NSF) operated under a cooperative agreement by Associated Universities, Inc. 
The Green Bank Interferometer is a facility of the NSF operated by the NRAO in support of NASA High Energy Astrophysics programs.
This research has made use of data from the UMRAO, which is supported by funds from the University of Michigan.


\appendix
\section*{FFA Opacities along the Plasma Sphere with the Isothermal King Model}
Here, we discuss the optical depth of free--free absorption in the plasma sphere, whose electron density follows the isothermal King model (\cite{King1972}).
The model is derived from the force balance between the pressure gradient and the self-gravity of the gas, assuming spherical symmetry and thermal equilibrium. 
The model gives the radial profile of the particle density, $\rho$, as a function of the radius, $r$, by
\begin{eqnarray}
\rho (r) = \rho_0 \left[ 1 + \left( \frac{r}{r_{\rm c}} \right)^2 \right]^{-\frac{3}{2}}.
\end{eqnarray}
Here, $\rho_0$ is the density at the center and $r_{\rm c}$ is the core radius.
If we assume the isothermal condition, the relation between the electron density, $n_{\rm e}$, the electron temperature, $T_{\rm e}$, and the FFA opacity, $\tau_0$, will be
\begin{eqnarray}
\tau_0 &=& 0.46 \int_{\rm LOS} T_{\rm e}^{-\frac{3}{2}} n_{\rm e}^2 dL \nonumber \\
       &=& 0.46 T_{\rm e}^{-\frac{3}{2}} \int_{\rm LOS} n_{\rm e}^2 dL \nonumber \\
       &=& 0.46 T_{\rm e}^{-\frac{3}{2}} n_{\rm e0}^2 \int_{\rm LOS} \left[ 1 + \left( \frac{r}{r_{\rm c}} \right)^2 \right]^{-3} dL. \label{eqn:ffainking}
\end{eqnarray}
Here, $n_{\rm e0}$ is the electron density at the center in cm$^{-3}$ and $L$ is the path length along the line of sight in pc.

\begin{figure}
  \begin{center}
    \FigureFile(78.88mm,72.64mm){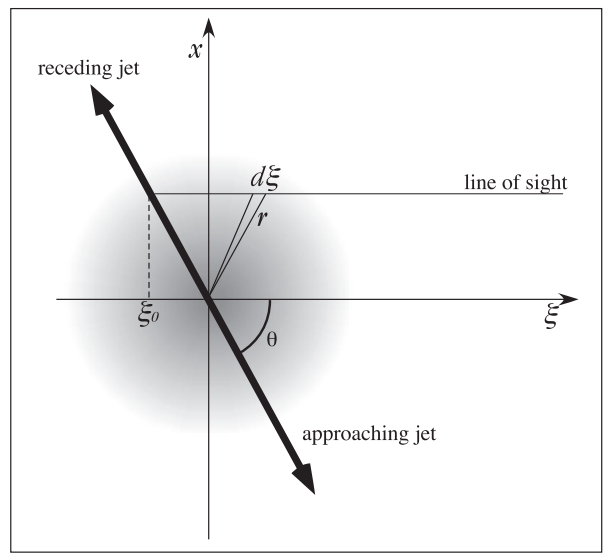}
  \end{center}
  \caption{FFA optical depth towards the twin jets along the plasma with the density profile of the isothermal King model.
Let the $xi$ and $x$ axes be parallel and perpendicular to the line of sight, respectively, and let the nucleus be the origin.
The jet, as a synchrotron emitter, is inclined from the line of sight by the viewing angle $\theta$.
The opacity integration, as a function of $x$, is taken from the jet at $z_0 = -\frac{x}{\tan \theta}$ to infinity.
}\label{fig:ambientmodel}
\end{figure}

\begin{figure}
  \begin{center}
    \FigureFile(78.54mm,63.92mm){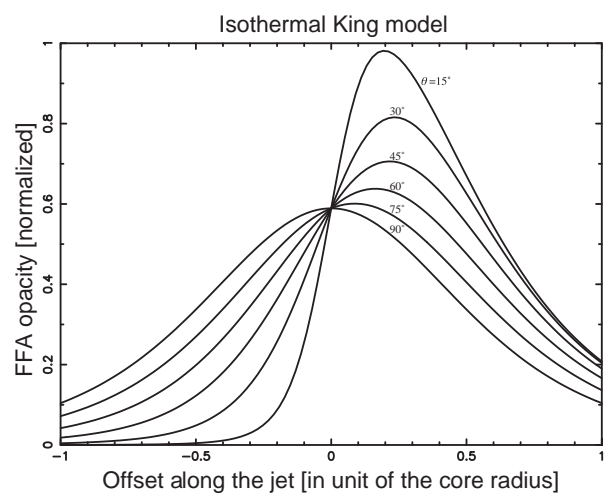}
  \end{center}
  \caption{Profiles of the FFA optical depth towards the jets with different viewing angles.
The horizontal axis is the offset along the jet in unit of the core radius, $r_{\rm c}$.
The negative offset stands for the approaching jet side. 
The vertical axis is the normalized opacity, when $T_{\rm e}^{-\frac{3}{2}} n_{\rm e0}^2 r_{\rm c}$ is unity.
}\label{fig:Kingmodel}
\end{figure}

Let us consider the configuration as shown in figure \ref{fig:ambientmodel}.
The integration of the optical depth in equation (\ref{eqn:ffainking}) should be taken from the jet at $\xi_0 = -\frac{x}{\tan \theta}$ to infinity.
The radius from the nucleus, $r$, is described in $(x, \xi)$ coordinate as $r^2 = x^2 + \xi^2$.
Therefore, equation (\ref{eqn:ffainking}) will be
\begin{eqnarray}
\tau_0 (x) &=& 0.46 T_{\rm e}^{-\frac{3}{2}} n_{\rm e0}^2 r_{\rm c}^6 \int_{\xi_0}^\infty (r_{\rm c}^2 + x^2 + \xi^2)^{-3} d\xi. \label{eqn:opacityxz}
\end{eqnarray}

Here, we use the mathematical formula
\begin{eqnarray}
\int (x^2 + C)^{-3} dx &=& \frac{x}{4C(x^2 + C)^2} + \frac{3x}{8C^2 (x^2 + C)} \nonumber \\
                       &+& \frac{3}{8C^2} \frac{1}{\sqrt{C}} \arctan \frac{x}{\sqrt{C}},
\end{eqnarray}
where $C > 0$ is a constant.
Equation (\ref{eqn:opacityxz}) will be replaced as
\begin{eqnarray}
\tau_0 (x) &=& 0.46 T_{\rm e}^{-\frac{3}{2}} n_{\rm e0}^2 r_{\rm c}^6
\left[
\frac{\xi_0}{4(r_{\rm c}^2 + x^2)(\xi_0^2 + r_{\rm c}^2 + x^2)^2} \right. \nonumber \\
           &+& \left. \frac{3\xi_0}{8(r_{\rm c}^2 + x^2)^2(\xi_0^2 + r_{\rm c}^2 + x^2)} \right. \nonumber \\
           &+& \left. \frac{3 \left( \frac{\pi}{2} - \arctan \frac{\xi_0}{\sqrt{r_{\rm c}^2 + x^2}} \right)}{8(r_{\rm c}^2 + x^2)^2 \sqrt{r_{\rm c}^2 + x^2}} \right]. \label{eqn:opacitiyinteg}
\end{eqnarray}
Let $x$ replaced with a dimensionless quantity, $x^{\prime} = x/r_{\rm c}$, then equation (\ref{eqn:opacitiyinteg}) reduces to 
\begin{eqnarray}
\tau_0 (x^{\prime}) &=& 0.46 T_{\rm e}^{-\frac{3}{2}} n_{\rm e0}^2 r_{\rm c}
\left[
\frac{x^{\prime}}{4\tan \theta (1 + x^{\prime 2})(1 + \left(\frac{x^{\prime}}{\sin \theta} \right)^2)^2} \right. \nonumber \\
  &+& \left. \frac{3 x^{\prime}}{8\tan \theta (1 + x^{\prime 2})^2 (1 + \left(\frac{x^{\prime} }{\sin \theta} \right)^2)} \right. \nonumber \\
  &+& \left. \frac{3\left(\frac{\pi}{2} - \arctan \frac{x^{\prime}}{\tan \theta \sqrt{1 + x^{\prime 2}}} \right)}{8(1 + x^{\prime 2})^2 \sqrt{1 + x^{\prime 2}}} \right]. \label{eqn:opacitfinal}
\end{eqnarray}

Figure \ref{fig:Kingmodel} shows opacity profiles along the jets with different viewing angles, calculated by equation (\ref{eqn:opacitfinal}).
As easily supposed, the slant jet axis causes an asymmetric profile of the opacity.
Since the electron density is larger towards the center, the opacity is larger within the core radius than outside.
When the radiation source is far enough from the center, the opacity is small at any given viewing angle.


\end{document}